\documentclass[twocolumn]{jpsj3}

\usepackage{graphicx,here,color,ulem}
\usepackage{amsmath, amssymb,bm}
\usepackage[dvipdfmx]{graphicx}

\title{Topological Properties of Magnetically Ordered Heavy-Fermion Systems\\ in the Presence of Mirror Symmetry}

\author{$^{1}$\name{Kazuhiro \name{Kimura}}\thanks{E-mail address: kimura.kazuhiro.85n@st.kyoto-u.ac.jp},$^{1,2}$\name{Tsuneya \name{Yoshida}}, and $^{1}$\name{Norio \name{Kawakami}} 
}
\inst{\address{$^{1}$Department of Physics, Kyoto University, Kyoto 606-8502, Japan}\\
$^{2}$Division of Physics, Tsukuba University, Tsukuba, Ibaraki 305-8571, Japan
}
\abst{We explore topological states with magnetic order in heavy-fermion systems by taking account of a mirror symmetry.
Although in the absence of spatial symmetry, there is no topological phase in the two-dimensional (2D) antiferromagnetic phases at half filling, we demonstrate that a topological phase emerges in the presence of mirror symmetry. 
This is explicitly shown for a 2D periodic Anderson model.
Furthermore, around quarter filling, our analysis shows that a half-metallic state emerges in the ferromagnetic phase, where a spin-selective gap opens, resulting in nontrivial properties characterized by a Chern number. 
In contrast to the previously proposed models, our scenario can even apply for spin-nonconserving systems in the presence of spin-orbit coupling.
}

\begin{document}
\maketitle
\section{Introduction}

Since the theoretical discovery of topological insulators (TIs) \cite{Kane_Mele_QSH_05,Kane_Mele_Z2_05}, topological materials have been the subject of intense theoretical and experimental investigation. Topological materials are characterized by their metallic surface (edge) states protected by the topology of the bulk wave function. A significant number of TIs have been found so far this decade, including topological crystalline insulators \cite{Liang_Fu_11,Hsieh_12,Tanaka_Ren_12,Dziawa_Kowalski_12}. Furthermore, closely related issues such as topological superconductors, \cite{Kitaev_01,Mourik_12,Rokhinson_12,Das_12,Yoshida_Sigrist_15,Daido_16} {and Weyl semimetals \cite{Murakami_07,Wan_11,Burkov_11,Weng_15,Xu_15,Lv_15} have also been the focus of intense interest, and topology has become a ubiquitous issue in condensed matter physics.

Recently, the notion of TIs has been extended to strongly correlated systems, such as topological Kondo insulators \cite{Dzero_10,Dzero_16}.
In particular, SmB$_6$ \cite{Neupane_13,Xu_14} has attracted much attention as a promising candidate for the topological Kondo insulator.
This compound has been known as a Kondo insulator for about 40 years, and it has been recently proposed that the saturation of its electrical resistivity at low temperatures is due to the topological surface current.
In addition to topological Kondo insulators, there are intriguing topological phenomena \cite{Hohenadler_11,Yoshida_Fujimoto_Kawakami_12,Tada_12,
Shitade_09,Chadov_10,Lin_10,Takimoto_11,Yan_12,Zhang_13,Lu_13,Hsieh_14,Kargarian_13,Weng_14}
 under electron correlations such as topological Mott insulators \cite{Pesin_Balents_10,Yoshida_Peters_Fujimoto_Kawakami_14,Yoshida_Kawakami_16,Wu_16}, interaction-reduced classifications \cite{Fidkowski_10,Fidkowski_11,Turner_11,Ryu_Zhang_12,Yao_Ryu_13,Qi_13,Lu_Vishwanath_12,Levin_12,Fidkowski_13,Gu_14,Hsieh_Chang_14,
Isobe_Fu_15,Yoshida_Furusaki_15,Wang_Potter_Senthil_14,You_Cenke_14,Wang_Senthil_14,Morimoto_15,Yoshida_Daido_17,Yoshida_Danshita_17}, and the competition between long-range-ordered phases and TIs \cite{Mong_10,Essin_Gurarie_12,Okamoto_14,Guo_11,He_11,Yoshida_13,He_12,Yoshida_SSTI_13,
Rachel_10,Varney_10,Yamaji_11,Zheng_11,Griset_12,Wu_12,Yu_11}.
The strong correlation effects on topological states are particularly interesting because of nontrivial properties originating from the interplay between topology and correlation.

In this paper, we focus on the topological properties in long-range-ordered phases such as ferromagnetic (FM) and antiferromagnetic (AFM) phases.
The emergence of topological properties in the magnetic phases has already been studied. For example, {an antiferromagnetic topological insulator (AFTI) \cite{Mong_10} has been proposed for three-dimensional (3D) systems. In such AFTIs, the translation symmetry is broken and the unit cell is doubled.
The AFM order breaks time-reversal ($\Theta$) and primitive-lattice translational ($T_{1/2}$) symmetries but preserves their combined symmetry $S=T_{1/2} \Theta$. Under this symmetry, the system has a $Z_2$ topological number, which is related to the strong index of 3D topological insulators. Note, however, that this $Z_2$ number is allowed only for 3D systems, as confirmed by the periodic table \cite{ten-fold_way_08,ten-fold_way_09,ten-fold_way_10,Slager_13,Slager_17,Shiozaki_nonsymm_16}. We here attempt to solve this problem and demonstrate that we can realize AFTIs even in two-dimensional (2D) systems by taking into account a mirror symmetry.

Concerning nontrivial properties in FM phases, a spin-selective topological insulator (SSTI) was proposed for heavy-fermion systems \cite{Yoshida_SSTI_13}. This is a half-metallic FM phase \cite{Beach_Assaad_08,Peters_12,Howczak_Spalek_12,Peters_Kawakami_12,Denis_Rok_13} around quar]ter filling, where one spin sector acquires a gap while the other remains gapless, which is called a spin-selective gap. If the insulating sector has a nontrivial topological number, we have an SSTI, which  is a unique TI embedded in a metallic phase.
Unfortunately, however, it has been shown that this phase emerges only in spin-conserving systems and is not generally compatible with the presence of spin-orbit coupling, which is necessary for  topological properties.
Here, we will overcome this difficulty and propose a way to realize an SSTI for a spin-nonconserving system in the presence of spin-orbit coupling.

We explore the above-mentioned topological states in 2D magnetically ordered phases by using an effective model of the topological Kondo insulator for heavy-fermion systems, and demonstrate how they can emerge by using the Hartree-Fock (HF) approximation.
Our main idea for realizing such topological states in 2D magnetic phases is to take into account crystal symmetries, in particular, a mirror symmetry. We address two kinds of magnetic phases: a 2D AFM phase at half filling and a half-metallic FM phase around quarter filling. By taking into account the mirror symmetry, we elucidate the remarkable facts that a 2D AFM phase can have a topologically nontrivial structure specified by a mirror Chern number and that a half-metallic FM phase can have a topologically nontrivial structure specified by a Chern number. An important point is that these states can appear for spin-nonconserving systems.

The rest of this paper is organized as follows. 
In Sect. \ref{sec:model}, we introduce the model and the method. We then
 discuss the results for an AFTI at half filling in Sect. \ref{sec:afm}, and a half-metallic FM topological insulator near quarter filling in Sect. \ref{sec:fm}. In Sect. \ref{sec:green}, we briefly address the electron correlation effect on our topological phases. A brief summary is given in Sect. \ref{sec:summary}.

\section{Model and Method \label{sec:model}}

We explore the topological properties of the heavy-fermion systems by employing a 2D periodic Anderson model with nonlocal $d$-$f$ hybridization \cite{Dzero_10,Dzero_16,Legner_Sigrist_14,Legner_Sigrist_15}.
Specifically, we analyze the following topological periodic Anderson model showing band inversion at X points \cite{Takimoto_11,Tran_12,Lu_13,Alexandrov_13}due to next-nearest-neighbor (n.n.n.) hopping, which is important for describing SmB$_6$.
Here, note that $d$-electrons are conduction electrons.
In addition to the strong interaction $U_f$ between $f$-electrons, we also consider the interaction $U_d$ between $d$-electrons.
The Hamiltonian reads 
\begin{subequations}
\label{eq:Hamiltonian}
\begin{eqnarray}
H&=&\sum_{\bm{k}}
\begin{pmatrix}
\bm{d}^{\dagger}_{\bm{k}}  &\bm{f}^{\dagger}_{\bm{k}}
\end{pmatrix}
\begin{pmatrix}
\epsilon^d_{\bm{k}}   &V_{\bm{k}}  \\
V^{\dagger}_{\bm{k}}&\epsilon^f_{\bm{k}} 
\end{pmatrix}
\begin{pmatrix}
\bm{d}_{\bm{k}}  \\
\bm{f}_{\bm{k}}
\end{pmatrix}
\nonumber \\
&+&\sum_{j; \alpha ( =d,f)}U_{\alpha}n^{\alpha}_{j\uparrow}n^{\alpha}_{j{\downarrow}}, 
\end{eqnarray}
with
\begin{eqnarray}
\epsilon^d_{\bm{k}} &=& [-2t_d(\cos{k_x}+\cos{k_y})\nonumber \\
&&-4t_d^{\prime}\cos{k_x}\cos{k_y} ]\sigma_0 ,   \\
\epsilon_{\bm{k}}^f &=& [ \epsilon_f-2t_f(\cos{k_x}+\cos{k_y})\nonumber \\
&&-4t_f^{\prime}\cos{k_x}\cos{k_y}]\sigma_0,  \\
V_{\bm{k}}&=& -2[ \sigma_x\sin{k_x} (V_1+V_2\cos{k_y}) \nonumber \\
&&+\sigma_y\sin{k_y} (V_1+V_2\cos{k_x})  ],    
\end{eqnarray}
\end{subequations}
where $\epsilon^d_{\bm{k}}$($\epsilon^f_{\bm{k}}$) is the dispersion of $d$-($f$-) electrons, $V_{\bm{k}}$ is a Fourier component of the nonlocal $d$-$f$ hybridization, and $\bm{k}$ is a wave number.
The annihilation operators are defined as
$\bm{d}_{\bm{k}}=
\begin{pmatrix}
d_{\bm{k}\uparrow}  &
d_{\bm{k}\downarrow}
\end{pmatrix}
^{T}$
 and 
$\bm{f}_{\bm{k}}=
\begin{pmatrix}
{f}_{\bm{k}\uparrow}  &
{f}_{\bm{k}\downarrow}
\end{pmatrix}
^{T}$.
The basis function for this model is 
$(d_{\uparrow}, d_{\downarrow}, f_{\uparrow}, f_{\downarrow})^T$,
where $\sigma_{i}$($i =0,x,y,z$) are the Pauli matrices for spins.
Here,
$t_d$, $t_f$, $t_d^{\prime}$, and $t_f^{\prime}$ are hopping parameters, $V_1$ and $V_2$ denote the $d$-$f$ hybridization, and
$\epsilon_f$ is the difference between the $d$- and $f$-electron energies.
We consider the above model on a 2D square lattice in the $x$-$y$ plane, which is a 2D version of the topological crystalline insulator in a 3D cubic lattice.
The system has inversion symmetry, so that the hybridization has odd parity, $V_{\bm{k}}=-V_{\bm{-k}}$.

In order to study the ground state of the model in Eq. (\ref{eq:Hamiltonian}), we employ the following HF approximation for the Coulomb term:
\begin{eqnarray}
n_{i \uparrow}^{\alpha}n_{i \downarrow}^{\alpha}
&\sim&n_{i \uparrow}^{\alpha} \langle n_{i \downarrow}^{\alpha} \rangle  + \langle n_{i \uparrow}^{\alpha} \rangle n_{i \downarrow}^{\alpha} -  \langle n_{i \uparrow}^{\alpha} \rangle \langle n_{i \downarrow}^{\alpha} \rangle, 
\end{eqnarray}
where $n_{i\sigma}^{\alpha}$($\alpha =d,f$) is the number operator.
Here, $\langle \cdots \rangle$ denotes the expectation value at zero temperature.

We introduce the mirror operation $M_z$, which inverts the $z$-axis,
\begin{eqnarray}
M_z&=&i\, \tau_z \otimes \sigma_z ,
\end{eqnarray}
where $\tau_i$($i=0,x,y,z$) specifies $d$- and $f$-electrons.
In order to consider the magnetically ordered topological insulating states with a mirror symmetry, we introduce the corresponding topological number.
First, recall that the Chern number in multiband systems is given as
\begin{eqnarray}
C  &=& \frac{1}{2\pi} \sum_{i }\int_{S} [\nabla_{\bm{k}}\times \mathcal{A}_i]_z dk_x dk_y,
\end{eqnarray}
where $\mathcal{A}_i(\bm{k})=-i\langle u_i(\bm{k})|\nabla_{\bm{k}}|u_i(\bm{k})\rangle$ is the $U(1)$ Berry connection,
 where $|u_i(\bm{k}) \rangle$ is a Bloch state with occupied band index $i$, which is an eigenstate of ${\cal H}(\bm{k})$.
In the mirror-symmetric system, all the eigenstates are characterized by their mirror parities and divided into two subspaces as
\begin{eqnarray}
\mathcal{H}(\bm{k})=
\begin{pmatrix}
{\mathcal H}_{ M_z=+i}(\bm{k}) &0 \\
0&{\mathcal H}_{M_z=-i}(\bm{k}) 
\end{pmatrix}
.
\end{eqnarray}
The net Chern number $C$ and the mirror Chern number $C_m$ are defined by the Chern numbers $C_{\pm i}$ obtained in each mirror subspace,
namely,
\begin{subequations}
\begin{eqnarray}
C&=&C_{M_z=+i}+C_{M_z=-i} ,\\
C_m&=&(C_{M_z=+i}-C_{M_z=-i})/2 .
\end{eqnarray}
\end{subequations}

\section{Results}

We here separately discuss the obtained results for the magnetic phases at half filling and around quarter filling separately.
The values of the parameters we employ in the following are 
$t_d=1$ (energy unit),  $t_d^{\prime}=-0.5, t_f=-t_d/5, t_f^{\prime}=-t_d^{\prime}/5, U_d=2$. Unless otherwise noted we set $(V_1 ,V_2)=(0.1,-0.4)$.
The choice of these parameters will be explained below. 
The magnetic properties of the system are studied by the HF method and the Chern number is calculated by the Fukui-Hatsugai method\cite{Fukui_Hatsugai_05}, which is efficient for numerical calculations.

\subsection{Antiferromagnetic phase\label{sec:afm}}

For a heavy-fermion system at half filling, there are some AFM phases and a Kondo insulating phase in the ordinary Doniac phase diagram. In contrast to a previous study \cite{Mong_10}, we here demonstrate that the AFTI can emerge in 2D mirror-symmetric systems.
The model we employ here is a topological mirror Kondo insulator introduced in Refs \citen{Legner_Sigrist_14,Legner_Sigrist_15}, and \citen{Rui-Xing_Zhang_16}, which is a mirror-symmetric extension of the topological Kondo insulator. This model was previously used to address a nonmagnetic Kondo insulating phase. The net Chern number is zero, $C=0$, because of the time-reversal symmetry, but there is still a possibility of having a non-zero mirror Chern number $C_m \neq 0$.
We elucidate below that the system can change from a paramagnetic phase to an AFM phase without breaking its mirror symmetry, thus leading to an AFTI. The AFM phase, where the magnetization is along the $z$-axis in our case, breaks a space translation symmetry $T_{1/2}$, and thus the period of the unit cell is doubled. Time-reversal symmetry is also broken by the magnetization, but we show that the net Chern number is zero by using the combined symmetry $S=\Theta T_{1/2}$ of the time-reversal $\Theta$ and primitive-lattice translation $T_{1/2}$. 
                           

We assume that the nesting vector is $\bm{Q}=(\pi,\pi)$ for the AFM phase, which is justified for $(t^{\prime}_d,t^{\prime}_f,V_2)=(0,0,0)$, see below.
The mean-field Hamiltonian is given by
\begin{subequations}
\begin{eqnarray}
{\cal H}^{mf}_{\bm{k}}&=&
\begin{pmatrix}
\epsilon^d_{\bm{k}} +h^d_{int} & V_{\bm{k}} \\
V_{\bm{k}}^{\dagger} &\epsilon^f_{\bm{k}} +h^f_{int}
\end{pmatrix}
, 
\end{eqnarray}
with
\begin{eqnarray}
\epsilon^d_{\bm{k}} &=& [-2t_d(\cos{k_x}+\cos{k_y})\eta_x \nonumber \\ 
& &-4t_d^{\prime}\cos{k_x}\cos{k_y}\eta_0 ]\sigma_0 , \\
\epsilon^f_{\bm{k}} &=& [ \epsilon_f \eta_0 -2t_f(\cos{k_x}+\cos{k_y})\eta_x   \nonumber \\ 
&&-4t_f^{\prime}\cos{k_x}\cos{k_y}\eta_0 ]\sigma_0 ,  \\
V_{\bm{k}}&=& -2 [ \sin{k_x}\cdot \sigma_x(V_1 \eta_x+V_2 \eta_0\cos{k_y})  \nonumber \\
&&+\sin{k_y}\cdot \sigma_y(V_1 \eta_x+V_2 \eta_0\cos{k_x}) ] ,          
\end{eqnarray}
\begin{eqnarray}
h^{\alpha}_{int}= U_{\alpha}
\begin{pmatrix}
\langle n^{A\alpha}_{0\downarrow} \rangle  &0 &0 &0 \\
0 &\langle n^{B\alpha}_{0\downarrow} \rangle &0 &0 \\
0 &0 &\langle n^{A\alpha}_{0\uparrow} \rangle \\
0 &0 &0 &\langle n^{B\alpha}_{0\uparrow} \rangle
\end{pmatrix}
,
\end{eqnarray}
\label{eqn:Hami}
\end{subequations}
where $\alpha=d,f$, and $\eta_{i}$($i=0,x,y,z$) are the Pauli matrices for sublattice indices. The basis function is
$(d_{\uparrow}^A,d_{\uparrow}^B, d_{\downarrow}^A,d_{\downarrow}^B, f_{\uparrow}^A, f_{\uparrow}^B,f_{\downarrow}^A, f_{\downarrow}^B)^T$.
The mirror operation in the sublattice is 
\begin{eqnarray}
M_z&=&i\, \tau_z \otimes \sigma_z \otimes \eta_0 .
\end{eqnarray}

\subsubsection{ Case of $(t^{\prime}_d,t^{\prime}_f,V_2)=(0,0,0)$ }
\begin{figure}[h!]
\centering
\includegraphics[width=\hsize,clip,bb=64 453 539 743]{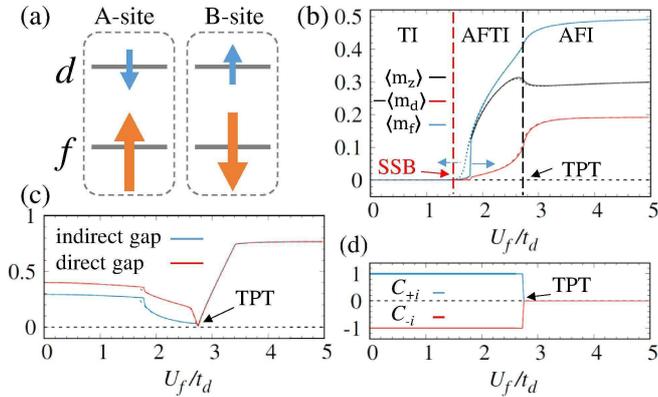}
\caption{
(Color) Magnetic and topological properties for a simplified model with only n.n. hopping and interaction, $(t_d^{\prime},t_f^{\prime},V_2)=(0,0,0)$ and $(U_d,V_1)=(2.0,0.1)$, obtained with the HF approximation: 
(a) spin configuration of the AFM phase, (b) staggered magnetic moments, (c) indirect gap and direct gap, (d) Chern number for each sector.
In (b), the red dashed line denotes a spontaneous symmetry breaking (SSB) transition, the black dashed line denotes a topological phase transition (TPT), and there is a small hysteresis loop because of the first-order transition.
In (c), the indirect gap is the band gap between the conduction and valence bands $\mathrm{min}(E^{\mathrm{conduction}}_{\bm{k}}-E^{\mathrm{valence}}_{\bm{k}^{\prime}})$ ($\bm{k}$ is not necessarily equal to $\bm{k}^{\prime}$) while the direct gap is the band gap at  wave number $\bm{k}$ $\mathrm{min}(E^{\mathrm{conduction}}_{\bm{k}}-E^{\mathrm{valence}}_{\bm{k}})$, where $\mathrm{indirect \, gap} \leq \mathrm{direct \, gap}$.
 We have a TI for $U_f<1.42$, an AFTI for $1.42<U_f<2.72$, and an AF trivial insulator (AFI) for $2.72<U_f$.
Except when $U_f=2.72$, the system is an insulator because of the finite indirect gap.
}
\label{fig:U0_all}
\end{figure}  

In order to see the essence of the results more clearly, we start with
a simplified model having only nearest-neighbor (n.n.) hopping and hybridization, i.e.,
$(t_d^{\prime},t_f^{\prime},V_2)=(0,0,0)$.
We first determine the easy axis of the magnetization by using second-order perturbation theory in the strong correlation limit. 
As a result, we conclude that the $z$-direction is the easy axis. 
The detail of the derivation is given in Appendix \ref{sec:spin}.
It turns out that the magnetic moments of $f$- and $d$-electrons align antiparallel at each site in the $z$-direction, as shown in Fig. \ref{fig:U0_all}(a).

The mean-field results for $(U_d,V_1)=(2,0.1)$ are summarized in Figs. \ref{fig:U0_all} (b)- \ref{fig:U0_all} (d). 
Note that the results are not sensitive to the value of $V_1$. 
We obtain an AFTI phase as shown in Fig. \ref{fig:U0_all}(b).
At $U_f=2.72$, there is a first-order magnetic phase transition, and the spin configuration for $U_f>2.72$ in Fig. \ref{fig:U0_all}(a), where $d$- and $f$-electrons align in the opposite directions, is in accordance with the second-order perturbation analysis.
We show the direct and indirect gaps in Fig. \ref{fig:U0_all}(c). 
The former is important for determining the topological structure, and we confirm that there is indeed a finite direct gap in the AFM phase.
Note that the phase transition in Fig. \ref{fig:U0_all}(c) is a Lifshitz transition, where gap closing seems to occur at a single point.
However, this is an accidental phenomenon due to our choice of parameters.
It is seen that the AFTI phase extends between $U_f=1.42$ and $U_f=2.72$, where the mirror Chern number takes a value of $C_m=1$ in Fig. \ref{fig:U0_all}(d) with the finite magnetization in Fig. \ref{fig:U0_all}(b).
In the strong interaction region,  there is a topological phase transition to a trivial phase at $U_f=2.72$, where both the direct and indirect gaps are closed. As seen from Fig. \ref{fig:U0_all}(d), the transition is accompanied by a change in the mirror Chern number from $C_m=1$ to $C_m=0$ while the net Chern number is zero.
This topological phase transition is triggered by the competition between the two types of gap. Namely, in the weak interaction region, the topologically nontrivial gap due to the nonlocal hybridization $V_1$ is dominant, while 
in the strong interaction region, the topologically trivial gap with the AFM order is dominant. The competition between the two different states gives rise to a topological phase transition accompanied by gap closing (see Appendix \ref{sec:direct}).

\subsubsection{ Case of $(t^{\prime}_d,t^{\prime}_f,V_2)=(-0.5,0.1,-0.4)$ }
We now investigate the model with a specific choice of the parameters, $(t_d^{\prime},t_f^{\prime},V_2)=(-0.5,0.1,-0.4)$. Importantly, these parameters can describe band inversions for the X points in the 3D Brillouin zone (see Appendix \ref{sec:three}), leading to a strong topological insulator phase, as observed for SmB$_6$ via angle-resolved photoemission spectroscopy measurements \cite{Neupane_13,Xu_14}. 
The results obtained for topological and magnetic properties at half filling are shown in Fig. \ref{fig:smb6_U2}(a).
A prominent feature in this model is that the system becomes metallic where the indirect gap is closed in the AFM phase even at half filling, as seen in Fig. \ref{fig:smb6_U2}(b). 
Note, however, that the topological properties still remain intact in this region because the direct gap is not closed. 
Namely, the Chern number is still well defined [Fig. \ref{fig:smb6_U2}(c)] in the region where the direct gap is open. 
Thus, the topological properties remain even in a ``metal'', and such a metal adiabatically connected to a topological insulator is called a topological semimetal. 
Note that this definition of a semimetal is standard in condensed matter but slightly different from that for Dirac/Weyl semimetals, which are zero-gap semiconductors by definition.
Finally, there are several topological phase transitions between different Chern numbers, as seen in Fig. \ref{fig:smb6_U2}(c).
This spin configuration is of the AFM-II type in Fig. \ref{fig:smb6_U2}(d).

Summarizing all these results, we arrive at the phase diagram shown in Fig. \ref{fig:phase-afm}.
The horizontal axis denotes the strength of the interaction $U_f$ and the vertical axis the strength of hybridization $V_1$. 
There are two AFM phases in Fig. \ref{fig:phase-afm}.
The above analysis for the dashed blue line in Fig. \ref{fig:phase-afm} ($V_1=0.1,V_2=-0.4$) also applies to the region $|V_1|<|V_2|$ where the spin configuration is of the AFM-II type.
In the AFM phase for these parameters, a semimetallic AFM topological phase is realized, which we refer to as an AFM topological semimetal.
The mirror Chern number has various values in the phase diagram, which is due to the presence of n.n.n. hopping and hybridization, and is enriched by $U_d$. 
The changes in the mirror Chern number are driven by the shift of the $f$-band. 
In general, a complex band structure brings about various topological numbers (mirror Chern numbers), for example, see Refs. \citen{Yoshida_Daido_17} and \citen{Rui-Xing_Zhang_16}.
For reference, in Appendix \ref{sec:afti}, we show the phase diagram for $U_d=0$, which is much simpler than the one discussed above.
\begin{figure}[h!]
\centering
\includegraphics[width=\hsize,clip,bb= 76 300 546 626]{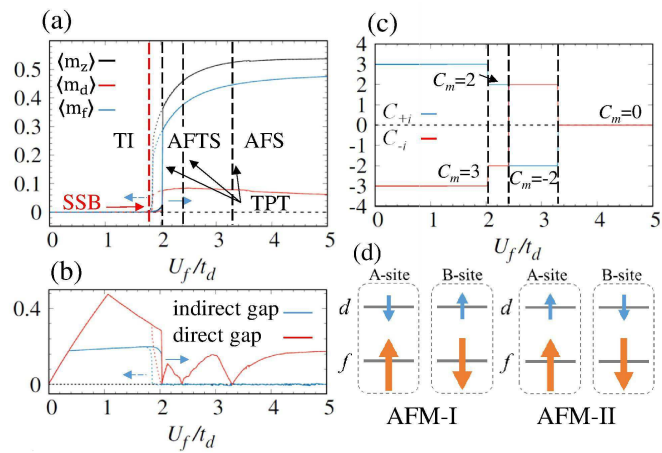}
\caption{
(Color) Magnetic and topological properties for the effective model of SmB$_6$, $(t_d^{\prime},t_f^{\prime},V_2)=(-0.5,0.1,-0.4)$ and $(U_d,V_1)=(2.0,0.1)$ at half filling:
(a) staggered magnetic moments, (b) indirect gap and direct gap, (c) Chern number for each sector, (d) spin configurations of the two AFM phases. 
In (a), there is a small hysteresis loop because of the first-order transition.
In (b), the region where the indirect gap is closed is semimetallic, and the points where the direct gap is closed denote the topological phase transitions.
In (c), there are two Chern numbers for two mirror sectors, and the change in the Chern numbers signals the topological phase transition.
We have an AF topological semimetal (AFSM) for $1.8<U_f<3.25$ and an AF trivial semimetal (AFS) for $U_f<3.25$.
}
\label{fig:smb6_U2}
\end{figure}                                                  
\begin{figure}[h!]
\centering
\includegraphics[width=0.8\hsize,clip,bb=87 368 458 660]{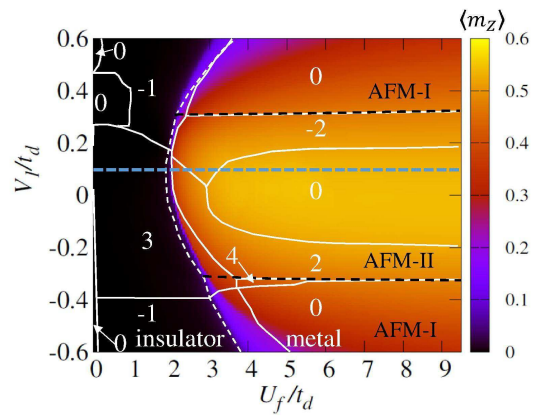}
\caption{
(Color) Phase diagram of mirror-symmetric AFTI at half filling for $U_d=2$ as a function of the interaction $U_f$ and  hybridization $V_1$.
The white (dashed) line denotes the topological (insulator-metal) phase transition line and
the black dashed line separates two AFM phases, i.e., AFM-I and AFM-II.
We set $V_2=-0.4$. 
In the metallic region, the indirect gap is closed; thus, there is a Fermi surface. However, the direct gap at the same wave 
number $\bm{k}$ is not closed; thus, the Chern number is still well defined.
The white numbers are mirror Chern numbers.
The mirror Chern numbers are 4, 3, 2, 0, -1, and -2 in this figure.
}
\label{fig:phase-afm}
\end{figure}                                                  

Here some comments on the difference between the current results and the previous ones are in order. So far, topological properties with the AFM order have been studied in Refs. \citen{Okamoto_14,Guo_11,He_11,He_12,Yoshida_13}, focusing on the systems with spin $U(1)$ symmetry. 
The Hamiltonian with spin $U(1)$ symmetry can be block-diagonalized for two spin sectors. 
In such a case, the topology of the AFM phase is characterized by the spin Chern number. 
In the presence of spin-orbit coupling, however, such $U(1)$ symmetry may disappear generally. Here, we stress that the AFTI in our analysis is more generic in the sense that our scenario does not require spin $U(1)$ symmetry. AFM systems respecting mirror symmetry with strong spin-orbit coupling are candidates for the AFTI proposed in this paper.
\subsection{Ferromagnetic phase\label{sec:fm}}

We now move on to an intriguing topological half-metallic state.
Around quarter filling in the Kondo lattice system, it has been known that a half-metallic FM phase dubbed a spin-selective Kondo insulator \cite{Beach_Assaad_08,Peters_12,Howczak_Spalek_12,Peters_Kawakami_12,Denis_Rok_13} appears, where a spin-selective gap opens, namely, one spin sector is metallic while the other is insulating. 
This has been demonstrated for spin-conserving systems and has been extended later to a topological version referred to as a spin-selective topological insulator (SSTI)  \cite{Yoshida_SSTI_13}, where the insulating sector has topologically nontrivial properties. 
A crucial problem in the previous proposals is that all the results on the SSTI rely on  spin $U(1)$ symmetry, which will disappear in the presence of spin-orbit coupling in general.  Thus, one might naively think that the SSTI cannot appear in reality.
To overcome this difficulty, we here demonstrate that by using a mirror symmetry, such a topological half-metallic state can indeed exist in the 2D FM phase.

\begin{figure}[t]
\centering
\includegraphics[width=\hsize,clip,bb=118 273 541 655]{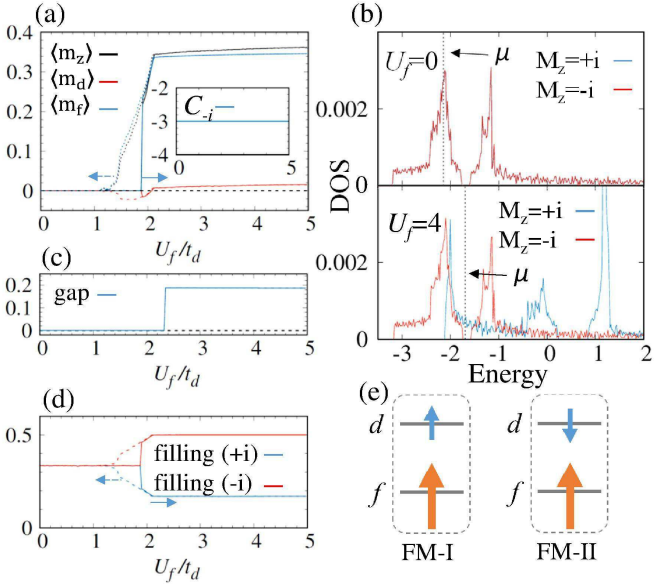}
\caption{
(Color) Magnetic and topological properties around quarter filling for $(U_d,U_f)=(0,4)$ and  filling of 0.335:
(a) magnetization of $f$- ($d$-) electrons, where 
the Chern number for the gapped sector is plotted in the inset. 
(b) DOS at $U_f=0$ and $U_f=4$, (c) gap of insulating sector, (d) electron filling for each mirror sector, (e) spin configurations. 
In (a), the red (blue) line represents $d$- ($f$-) electron magnetization and the black line the total magnetization.
For all $U_f$, the system has the same Chern number $C_{-i}=-3$ in the  mirror sector $M_z=-i$. 
In (b), we show the DOS for the sector of $M_z=+i$ ($M_z=-i$) by the blue (red) line,
where the chemical potential $\mu$ is indicated by the dashed black line.
In (c), the blue line is the gap of the mirror sector $M_z=-i$. 
In the region of $U_f>2.3$, the sector $M_z=-i$ is an insulator. 
We have a metallic state at $U_f=0$, while at $U_f=4$, we have the SSTI, where the sector $M_z=+i$ ($M_z=-i$) is a metal (an insulator).
}
\label{fig:smb6_U0}
\end{figure}                                                  
\begin{figure}[t]
\centering
\includegraphics[width=0.8\hsize,clip,bb=110 350 481 645]{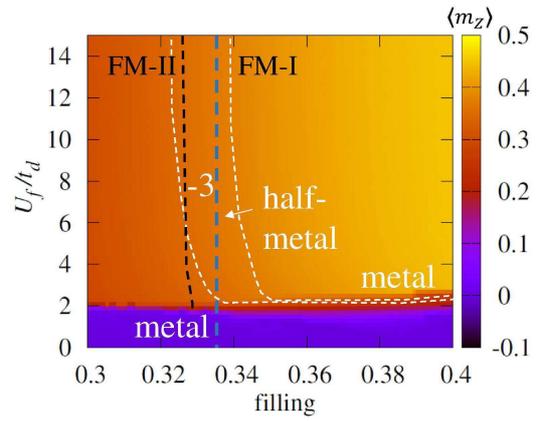}
\caption{
(Color) Phase diagram of mirror-selective topological insulator around quarter filling for $U_d=0$.
The white dashed line denotes the insulator-metal phase transition line and
the black dashed line separates two FM phases, i.e.,  FM-I and FM-II.
In the half-metallic region, the sector $M_z=-i$ has a finite gap and the sector $M_z=+i$ is metallic.
The white numbers are mirror Chern numbers of the sector $M_z=-i$ in the whole  region. 
The blue dashed line represents the filling of 0.335.
}
\label{fig:phase-fm}
\end{figure}                                                  

In Figs. \ref{fig:smb6_U0} and \ref{fig:phase-fm}, we show the results obtained around quarter filling.  
At a filling of 0.335 and $U_d=0$, a FM phase emerges, as seen in Fig. \ref{fig:smb6_U0}(a), where the magnetization has a hysteresis loop.
From the density of states (DOS) shown in Fig. \ref{fig:smb6_U0}(b), we find that the system is metallic at $U_f=0$, whereas the system is half-metallic at $U_f=4$  with the $M_z=+i$ sector being metallic while the $M_z=-i$ sector is insulating, as seen in Fig. \ref{fig:smb6_U0}(c).
This mirror-selective gap gives rise to the nontrivial topological number $C_{-i}=-3$ in Fig. \ref{fig:smb6_U0}(a), resulting in a mirror-selective topological insulator where the filling of the insulating sector is always half, as seen in Fig. \ref{fig:smb6_U0}(d).
This spin configuration in Fig. \ref{fig:smb6_U0}(e) is of the FM-I type.

All these results are put together in the phase diagram of Fig. \ref{fig:phase-fm}, shown as functions of the strength of the interaction $U_f$ and the filling in the system. 
There are two FM phases having different types of spin configuration in Fig. \ref{fig:smb6_U0}(e) and the system has competition between two magnetic orders. We also study the case including the finite interaction $U_d$, as shown in Appendix \ref{sec:msti}. At $U_d=2$, there is no topological phase, in  contrast to the above-mentioned case of $U_d=0$ in Fig. \ref{fig:phase-fm}, which shows a nontrivial topological phase in some parameter region.
Summarizing, we find the mirror-selective topological insulator in a half-metallic FM phase, which can emerge for spin nonconserving systems, in contrast to the previous proposals.

\subsection{Electron correlation effect\label{sec:green}}
So far, we have discussed the nontrivial topological states in the AFM phase and the half-metallic FM phase in the HF approximation.
One may ask what will happen if electron correlations are taken into account beyond the HF treatment.
Here, we argue that the topological properties obtained from the mean-field Hamiltonian can persist even if we consider electron correlations by, for example, dynamical mean-field theory, provided the Mott transition is absent according to Refs. \citen{Volovik_03,Gurarie_11,Zhong_Wang_12,Essin_Gurarie_11,Zhong_13}.
Recall that the Chern number of each mirror sector is given in terms of the Green's function as
\begin{eqnarray}
C_{\sigma}&=& \int \frac{d \omega d^2k}{24 \pi^2}{\rm Tr} [ \epsilon^{\mu \nu \rho}\, G_{\sigma}\partial_{\mu} G_{\sigma}^{-1}\, G_{\sigma}\partial_{\nu} G_{\sigma}^{-1}\, G_{\sigma}\partial_{\rho} G_{\sigma}^{-1}], \nonumber \\
\end{eqnarray}
where $\epsilon_{\mu \nu \rho}$ is a totally antisymmetric Levi-Civita tensor, and $(\partial_0,\partial_1,\partial_2)=(\partial_{\omega},\partial_{k_x},\partial_{k_y})$, $k=(\omega, \bm{k})$. Summation is assumed over repeated indices $\mu, \nu, \rho=0,1,2$.
 $\sigma$ specifies the mirror parity and $G_{\sigma}$ is the full single-particle Green's function, which is related to the free Green's function $G_{\sigma 0}$ via $G^{-1}_{\sigma}(i\omega ,\bm{k})=G^{-1}_{\sigma 0}(i\omega ,\bm{k})-\Sigma _{\sigma}(i\omega ,\bm{k})$, where $\Sigma _{\sigma}(i\omega ,\bm{k})$ is the self-energy.
In the present treatment, $G_{\sigma}$ is a $4 \times 4$ ($2 \times 2$) matrix in the AFM (half-metallic FM) case.
According to Refs. \citen{Zhong_Wang_12} and \citen{Zhong_13}, the Chern number is determined by the topological Hamiltonian $h^{\rm eff}_{\sigma}(\bm{k})=-G^{-1}_{\sigma}(0,\bm{k})=-G^{-1}_{\sigma 0}(0,\bm{k})+\Sigma_{\sigma} (0,\bm{k})$.
This is because the Chern number does not change under the smooth deformation as
\begin{eqnarray}
G_{\sigma}(i\omega, \bm{k}, \lambda)=(1-\lambda)G_{\sigma}(i\omega, \bm{k})+\lambda [i\omega+G^{-1}_{\sigma}(0,\bm{k})]^{-1}, \nonumber \\
\end{eqnarray}
where $\lambda \in [0,1]$, provided  ${\rm det}G_{\sigma} \neq 0$ and ${\rm det}G_{\sigma}^{-1} \neq 0$ are satisfied.
The cases of ${\rm det}G_{\sigma} = 0$ and ${\rm det}G_{\sigma}^{-1} = 0$ respectively correspond to the gap closing
or the emergence of Mott insulators with ${\rm Im}[\Sigma (0,\bm{k})]\rightarrow -\infty$.
Therefore, provided the Mott transition does not occur, the electron correlation effect on the AFM (half-metallic FM) phase can be treated with the renormalized band insulator, and thus the HF results may not be changed qualitatively, although the phase diagram should be modified quantitatively.

\section{Summary\label{sec:summary}}
We have explored two topological states in the AFM/FM phases by taking account of the mirror symmetry in heavy-fermion systems.
Concretely, in reference to topological crystalline insulators, we have proposed 2D topological crystalline insulating states in magnetic phases for interacting systems.
In particular, we have shown that in the AFM phase at half filling there is a topological state characterized by a mirror Chern number. 
In the case of a SmB$_6$ film, an AFM topological semimetallic phase is expected.
We have also shown that in the half-metallic FM phase around quarter filling, the spin-selective topological insulating state characterized by a Chern number is realized.


In contrast to the previous studies, which assumed spin $U(1)$ symmetry to obtain such topological properties in the magnetic phases, our proposal is that these phases can be realized even in the absence of spin $U(1)$ symmetry by taking into account crystalline symmetry in magnetic phases. Generally, spin $U(1)$ symmetry is not preserved in the presence of spin-orbit coupling; thus, the present scenario without respecting spin $U(1)$ symmetry will provide a feasible platform to realize magnetic topological insulators for 2D systems.

In this paper, we have employed the HF approximation to address the above phases. We have discussed the correlation effects qualitatively and shown that the topological properties of these states may not change in the presence of correlation effects. Nevertheless, more elaborate calculations should be carried out to confirm this conclusion, which is now under consideration.
In addition, a 3D version of the mirror-selective topological insulator has been discussed \cite{Peters_18}. It might be interesting to study how our mirror-selective topological insulator extends to three dimensions by increasing the thickness of the layers.


\section*{Acknowledgments}
We thank R. Peters for the fruitful discussion.
This work was partly supported by JSPS KAKENHI Grant No. JP15H05855 and No. JP16K05501.
The numerical calculations were performed on the supercomputer at the Institute for Solid State Physics in the University of Tokyo, and SR16000 at Yukawa Institute for Theoretical Physics in Kyoto University.


\appendix
\section{ Spin Configurations in Strong Correlation Limit \label{sec:spin}}
In this Appendix, we determine the magnetization axis by using second-order perturbation theory to clarify whether the magnetization is mirror-symmetric or not. For simplicity, we use the model with $V_2=0, \, t_d^{\prime}=0, \, t_f^{\prime}=0$ at half filling. 
Starting from the strong correlation limit $U_f \gg 1$, we use the normalized hybridization $V_1/U_f$ as a perturbation parameter.
We use the following relations between the spins $S^{\alpha}_i$ ($\alpha =d,f$ and $i=x,y,z$) of $d$-/$f$-electrons 
and the fermionic operators $d_{i{\sigma}}$ and $f_{j\sigma}$ with spin indices $\sigma=\uparrow ,\downarrow$ and site indices $i$, $j$:
\begin{subequations}
\begin{eqnarray}
S_{xi}^d \cdot S_{xj}^f &=&\frac{1}{2}\sum_{\sigma=\uparrow ,\downarrow} \left( d_{i\sigma}^{\dagger}d_{i\bar{\sigma}}f_{j\sigma}^{\dagger}f_{j\bar{\sigma}}+d_{i\sigma}^{\dagger}d_{i\bar {\sigma}}f_{j\bar{\sigma}}^{\dagger}f_{j\sigma} \right),\nonumber \\
\\
S_{yi}^d \cdot S_{yj}^f &=&\frac{1}{2}\sum_{\sigma=\uparrow ,\downarrow} \left( - d_{i\sigma}^{\dagger}d_{i\bar {\sigma}}f_{j\sigma}^{\dagger}f_{j\bar{\sigma}}+d_{i\sigma}^{\dagger}d_{i\bar {\sigma}}f_{j\bar{\sigma}}^{\dagger}f_{j\sigma}\right) ,\nonumber \\
\\
S_{zi}^d \cdot S_{zj}^f &=& \frac{1}{2}\sum_{\sigma=\uparrow ,\downarrow} \left(d_{i\sigma}^{\dagger}d_{i\sigma}f_{j\sigma}^{\dagger}f_{j\sigma}-d_{i\sigma}^{\dagger}d_{i\sigma}f_{j\bar{\sigma}}^{\dagger}f_{j\bar{\sigma}} \right) ,\nonumber \\
\\
n_{i}^d \cdot n_{j}^f &=& \sum_{\sigma=\uparrow ,\downarrow} \left( d_{i\sigma}^{\dagger}d_{i\sigma}f_{j\sigma}^{\dagger}f_{j\sigma}+d_{i\sigma}^{\dagger}d_{i\sigma}f_{j\bar{\sigma}}^{\dagger}f_{j\bar{\sigma}}\right). \nonumber \\
\end{eqnarray}
\end{subequations}
The hybridization term then results in the following exchange interaction via the 
second-order perturbation:
\begin{eqnarray}
H^{\prime}&=&\sum_{i\in x} J_{df}^{ix} S_{xi}^d\cdot S_{xi-1}^f + J_{df}^{iy} S_{yi}^d\cdot S_{yi-1}^f+ J_{df}^{iz} S_{zi}^d\cdot S_{zi-1}^f \nonumber \\
&+&\sum_{i\in x} J_{df}^{ix} S_{xi-1}^d\cdot S_{xi}^f + J_{df}^{iy} S_{yi-1}^d\cdot S_{yi}^f+ J_{df}^{iz} S_{zi-1}^d\cdot S_{zi}^f \nonumber \\&+&\sum_{j\in y} J_{df}^{jx} S_{xj}^d\cdot S_{xj-1}^f + J_{df}^{jy} S_{yj}^d\cdot S_{yj-1}^f+ J_{df}^{jz} S_{zj}^d\cdot S_{zj-1}^f \nonumber\\ &+&\sum_{j\in y} J_{df}^{jx} S_{xj-1}^d\cdot S_{xj}^f + J_{df}^{jy} S_{yj-1}^d\cdot S_{yj}^f+ J_{df}^{jz} S_{zj-1}^d\cdot S_{zj}^f,\nonumber \\
\end{eqnarray}
where $J^{i,\alpha}_{df}$ ($i=x,y$ and $\alpha=x,y,z$) are the coupling constants between $d$- and $f$-electrons.
For the spin configuration along the $x$-axis, the coupling constants satisfy $J^{ix}_{df}=J_{df}>0, J^{iy}_{df}=-J_{df}<0$, and $J^{iz}_{df}=-J_{df}<0$, while for the $y$-axis, they satisfy $J^{jx}_{df}=-J_{df}<0, J^{jy}_{df}=J_{df}>0, J^{jz}_{df}=-J_{df}<0$ (see  Table. A$\cdot$1), and $J_{df}=2V_1^2 (\frac{1}{\epsilon_f}+\frac{1}{U_f-\epsilon_f})>0$, where $\epsilon_f$ is the chemical potential of $f$-electrons.
From only this constraint, we cannot yet determine the spin configuration at the ground state.
When $t_{dd}=t_{ff}=0$, the spins can be polarized along the $x$- or $y$-direction without energy loss.

We then consider another perturbation expansion in $t_f/U_f$ for $U_f \gg 1$.
It induces the AFM interaction $J_{ff}\bm{S}^f_i\cdot \bm{S}^f_j$($J_{ff}<0$), 
giving rise to frustration in these cases.
As a result, the easy axis of the magnetization is the $z$-axis, preserving the mirror symmetry, and the ground state of this model prefers the configuration having the staggered AFM order in Fig. \ref{fig:config_z}, where all other magnetic configurations are frustrated. 
Those for $d$-electrons and $f$-electrons align in the opposite directions at the same site.
\begin{table}[htb]
  \begin{center}
\caption{
Exchange couplings obtained for the spin-half 2D periodic Anderson model with nonlocal $d$-$f$ hybridization using second-order perturbation theory from the strong correlation limit. 
$J^x_{df}, J^y_{df}$, and $ J^z_{df}$ are coupling constants for the effective spin model, where superscripts $x$, $y$, and $z$ specify the quantization axis for the spin. 
There are two spin configurations along the $x$- and $y$-axes.
$+$ ($-$) means an antiferromagnetic (AF) [ferromagnetic (F)] coupling.
}
\begin{tabular}{|l|c|c|c|} \hline
       & $J^x_{df}$ & $J^y_{df}$ & $J^z_{df}$ \\ \hline \hline
     $x$-axis & $+$(AF) & $-$(F) & $-$(F) \\
     $y$-axis & $-$(F) & $+$(AF) & $-$(AF)  \\ \hline
\end{tabular}
\end{center}
\label{tab:coupling}  
\end{table}
\begin{figure}[H]
\centering        
\includegraphics[width=0.8\hsize,clip,bb= 40 497 358 740]{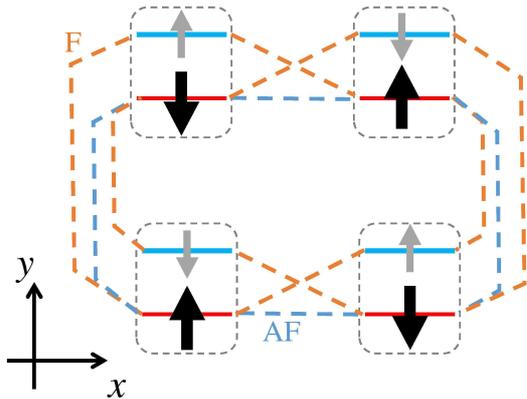}
\caption{
(Color) Spin configuration in real space, where all the spins align along the $z$-axis.
The blue (red) lines denote the $d$- ($f$-) level and the bold black/gray arrows denote spins.  
The dashed blue (orange) lines indicate the AF (F) coupling, where we consider the perturbation in $V_1/U_f$ and $t_f/U_f$.
There is no frustration and a configuration having a staggered AFM phase is realized, where 
$f$- and $d$-electrons align antiparallel at each site.
}
\label{fig:config_z}  
\end{figure}
We further consider the perturbation expansion in $t_d/U_d$ for $U_d\gg 1$ because $d$-electrons have reasonably strong correlation.
It turns out that this term induces the AFM interaction $J_{dd}\bm{S}^d_i\cdot \bm{S}^d_j$($J_{dd}<0$) with no frustration in this order if the easy axis of magnetization is the $z$-axis.
In this situation, the coupling constants are modified as $J_{df}=2V_1^2 (\frac{1}{U_d+\epsilon_f}+\frac{1}{U_f-\epsilon_f})>0$ but the sign is unchanged.
We thus conclude that the easy axis of the magnetization is the $z$-axis and a staggered AFM phase, where $f$- and $d$-electrons align antiparallel at each site, is realized.
\section{ Competition Between Two Types of Gap in the Case of $(t^{\prime}_d, t^{\prime}_f, V_2)=(0,0,0)$ \label{sec:direct}}

In this Appendix, we investigate the competition between two types of gap, topologically trivial and  nontrivial gaps, in a simplified model with only n.n. hopping and interaction, $(t^{\prime}_d,t^{\prime}_f,V_2)=(0,0,0)$ and $(U_d,V_1)=(2.0,0.1)$. There are two origins of this competition, one is the shift of the $f$-band $U_f \langle n_{0\sigma}^{AorB\alpha}\rangle$ and the other is the change in the nonlocal hybridization $V_{\bm{k}}$.

We first note that the Hamiltonian [Eq. (\ref{eqn:Hami})] can be block-diagonalized for two mirror sectors in the presence of mirror symmetry. We focus on one of the mirror sectors because both have the same band structure. The sector $M_z=+i$ is constructed from ($d_{\uparrow}^A, d_{\uparrow}^B, f_{\downarrow}^A, f_{\downarrow}^B$) while the sector $M_z=-i$ is constructed from ($d_{\downarrow}^A, d_{\downarrow}^B, f_{\uparrow}^A, f_{\uparrow}^B$). At $U_f=0$, a band inversion occurs between the second and third bands from the bottom in Fig. \ref{fig:bands}(a). Here, for simplicity, four bands are labeled as follows; the first and fourth bands from the bottom originate from $d$-electrons, whereas the second and third bands originate from $f$-electrons. The insulating phases are classified according to the band structure; (i) a band inversion occurs between the second and third bands, and the chemical potential lies between these $d$- and $f$-bands, (ii) a band inversion does not occur and the chemical potential lies between the $d$- and $f$-bands, and (iii) the chemical potential lies between different $d$-bands. The first type (i) has the nontrivial topological structure shown in Figs. \ref{fig:bands}(a)- \ref{fig:bands}(c), whereas the second type (ii) in Fig. \ref{fig:bands}(d) and the third type (iii) in Fig. \ref{fig:bands}(e) do not have nontrivial topological structures.

\begin{figure}[H]
\centering        
\includegraphics[width=0.5\hsize,clip,bb= 50 50 220 390]{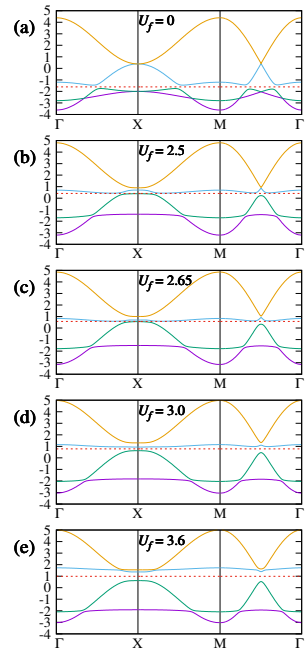}
\caption{
(Color) Energy spectra for a simplified model with only n.n. hopping and interaction, $(t^{\prime}_d,t^{\prime}_f,V_2)=(0,0,0)$ and $(U_d,V_1)=(2.0,0.1)$, obtained with the HF approximation: (a) $U_f=0$, (b) $U_f=2.5$, (c) $U_f=2.65$, (d) $U_f=3.0$, (e) $U_f=3.6$.
The red dashed line denotes the position of the chemical potential.
}
\label{fig:bands}  
\end{figure}

We first discuss the shift of the $f$-band, which is the Hartree shift due to the interaction $U_f$. To this end, it is sufficient to consider the Hartree shift in sublattices A and B only for one of the mirror sectors. The linear dependence of the direct gap on $U_f$ after the TPT in Fig. \ref{fig:U0_all}(c) can be understood by this Hartree shift, as described below. Before the TPT, the chemical potential is between the inverted $d$- and $f$-bands but after the TPT, the chemical potential is between the non-inverted $d$- and $f$-bands or between the $d$-bands. In the region where the chemical potential lies between the inverted $d$- and $f$-bands, the direct gap proportional to the Hartree shift for one of the $f$-bands causes the linear dependence on $U_f$ as seen in Fig. \ref{fig:U0_all}(c). On the other hand, when the chemical potential lies between different $d$-bands, the direct gap is not affected by the Hartree shift for the $f$-band, leading to an almost unchanged gap size. Since the direct gap caused by the Hartree shift does not have a topologically nontrivial structure, the system is topologically trivial. 

\begin{figure}[H]
\centering        
\includegraphics[width=0.5\hsize,clip,bb= 50 50 220 390]{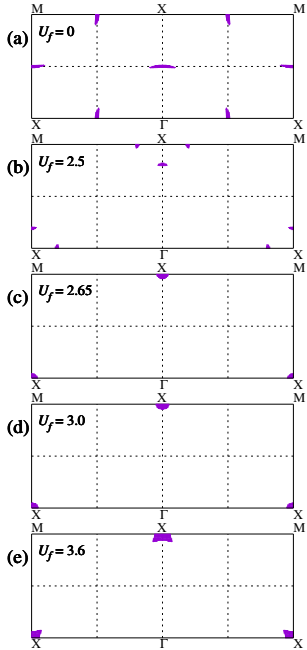}
\caption{
(Color) Wave number that brings about the lowest direct gap for a simplified model with only n.n. hopping and interaction, $(t^{\prime}_d,t^{\prime}_f,V_2)=(0,0,0)$ and $(U_d,V_1)=(2.0,0.1)$, obtained with the HF approximation: (a) $U_f=0$, (b) $U_f=2.5$, (c) $U_f=2.65$, (d) $U_f=3.0$, (e) $U_f=3.6$.
}
\label{fig:direct}  
\end{figure}

The above explanation becomes much clearer if we take into account the effect of the nonlocal hybridization $V_{\bm{k}}$, which is an odd function of $\bm{k}$. An important point is that $V_{\bm{k}}$ vanishes at the X points ($V_{\bm{k}=\rm{X}}=0$), which allows us to understand the characteristic behavior, in particular, in the region before the TPT in Fig. \ref{fig:U0_all}(c). By calculating the wave number dependence of the gap size, we determine the wave number that gives the smallest direct gap in Fig. \ref{fig:direct}. At $U_f=0$, this wave number is located in the middle between the $\Gamma$ and X points. With increasing $U_f$, the band structure around the Fermi level is changed by the Hartree shift as in Figs. \ref{fig:bands}(b), and \ref{fig:bands}(c) and the above-mentioned wave number approaches the X point in Fig. \ref{fig:direct}, making the gap size smaller. This is because around the X point the hybridization $V_{\bm{k}}$ becomes small, and therefore the direct gap rapidly decreases around the TPT in Fig. \ref{fig:U0_all}(c). Just after the TPT, the above wave number is still located around the X point in Figs. \ref{fig:direct}(d) and \ref{fig:direct}(e). Thus, the direct gap is caused only by the Hartree shift, giving rise to a linear dependence on $U_f$, as already mentioned above.

\section{ Three-Dimensional Effective Model of SmB$_6$ \label{sec:three}}
The 3D effective model Hamiltonian \cite{Legner_Sigrist_14,Legner_Sigrist_15,Rui-Xing_Zhang_16} reads
\begin{subequations}
\begin{eqnarray}
H_{\bm{k}}=\sum_{\bm{k}}
\begin{pmatrix}
\bm{d}^{\dagger}_{\bm{k}}  &\bm{f}^{\dagger}_{\bm{k}}
\end{pmatrix}
\begin{pmatrix}
\epsilon^d_{\bm{k}}   &V_{\bm{k}}  \\
V^{\dagger}_{\bm{k}}&\epsilon^f_{\bm{k}} 
\end{pmatrix}
\begin{pmatrix}
\bm{d}_{\bm{k}}  \\
\bm{f}_{\bm{k}}
\end{pmatrix}
\end{eqnarray}
with
\begin{eqnarray}
\epsilon^d_{\bm{k}} &=& [-2t_d(\cos{k_x}+\cos{k_y}+\cos{k_z})\nonumber \\
&&-4t_d^{\prime}(\cos{k_x}\cos{k_y}+\cos{k_y}\cos{k_z}
\nonumber \\
&&+\cos{k_z}\cos{k_x})]\sigma_0 ,  \nonumber \\
\\
\epsilon_{\bm{k}}^f &=& [ \epsilon_f-2t_f(\cos{k_x}+\cos{k_y}+\cos{k_z})\nonumber \\
&&-4t_f^{\prime}(\cos{k_x}\cos{k_y}+\cos{k_y}\cos{k_z}
\nonumber \\
&&+\cos{k_z}\cos{k_x})]\sigma_0, \nonumber \\ \\
V_{\bm{k}}&=& -2[\sigma_x \sin{k_x}(V_1+V_2(\cos{k_y}+\cos{k_z}))\nonumber \\
&&+\sigma_y \sin{k_y}(V_1+V_2(\cos{k_z}+\cos{k_x}))\nonumber \\
&&+\sigma_z \sin{k_z}(V_1+V_2(\cos{k_x}+\cos{k_y}))],\nonumber \\ 
\end{eqnarray}
\end{subequations}
where $\epsilon^d_{\bm{k}}$($\epsilon^f_{\bm{k}}$) is the dispersion of $d$-($f$-) electrons, $V_{\bm{k}}$ is the Fourier component of the nonlocal $d$-$f$ hybridization, and $\bm{k}$ is the wave number.
The annihilation operators are defined as
$\bm{d}_{\bm{k}}=
\begin{pmatrix}
d_{\bm{k}\uparrow}  &
d_{\bm{k}\downarrow}
\end{pmatrix}
^{T},$
$\bm{f}_{\bm{k}}=
\begin{pmatrix}
{f}_{\bm{k}\uparrow}  &
{f}_{\bm{k}\downarrow}
\end{pmatrix}
^{T}$.
The basis function for this model is 
$(d_{\uparrow}, d_{\downarrow}, f_{\uparrow}, f_{\downarrow})^T$,
where $\sigma_{i}$($i =0,x,y,z$) are the Pauli matrices for spins.
The concrete values of the parameters we employ are 
$t_d=1$ (energy unit),  $t_d^{\prime}=-0.5, t_f=-t_d/5, t_f^{\prime}=-t_d^{\prime}/5$. Note that this model has three band inversions at X points \cite{Takimoto_11,Tran_12,Lu_13,Alexandrov_13} as shown in  Fig. \ref{fig:band}. Our 2D effective model is introduced to properly take into account this inversion property.
\begin{figure}[H]
\centering        
\includegraphics[width=0.8\hsize,clip,bb=84 59 311 203]{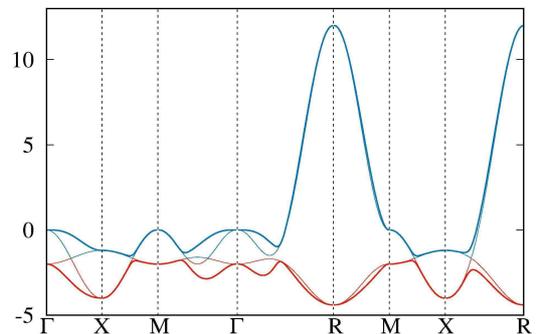}
\caption{
(Color) Energy spectrum of the model for 3D SmB$_6$.
The thick lines are the band structures of $d$- and $f$-electrons, $(V_1,V_2)=(0.1,-0.4)$.
The thin lines show the bare energies of $d$- and $f$-bands for the same parameters but with vanishing hybridization $(V_1,V_2)=(0,0)$. The band inversion at the X point is realized.
}
\label{fig:band}  
\end{figure}
\section{Antiferromagnetic Topological Insulator at Half Filling for  $U_d=0$\label{sec:afti}}
We investigate the topological and magnetic properties at half filling 
in the case of  $U_d=0$. The obtained phase diagram is shown in
Fig. \ref{fig:pd-afm-U0}. In contrast to the case with finite  $U_d$ shown in Fig. \ref{fig:phase-afm}, the phase diagram is much simpler, but we still find
a topologically nontrivial region. Note that in the AFM phase, the system becomes a topological semimetal, as mentioned in the main text.
\begin{figure}[H]
\centering        
\includegraphics[width=0.8\hsize,clip,bb=76 453 446 751]{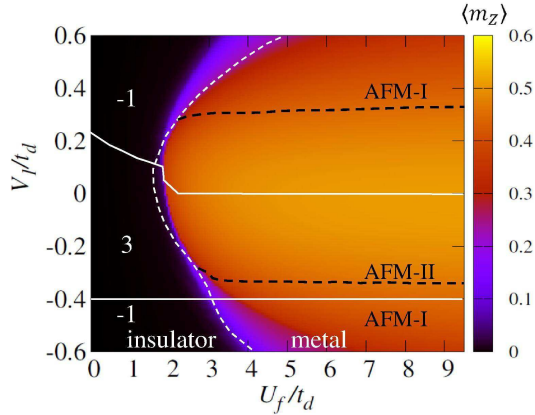}
\caption{
(Color) Phase diagram of mirror-symmetric AFTI at half filling for $U_d=0$. The white (dashed) line denotes the topological (insulator-metal) phase transition line. Here, $V_1$ is the n.n. hybridization and the n.n.n. hybridization $V_2=-0.4$ is fixed.
The white numbers are mirror Chern numbers.
In the metallic region, the indirect gap is closed, as mentioned in the main text. 
}
\label{fig:pd-afm-U0}  
\end{figure}

\section{Mirror-Selective Topological Insulator near Quarter Filling for $U_d=2$\label{sec:msti}}
We investigate topological and magnetic properties around quarter filling 
in the case of  $U_d=2$. The obtained phase diagram is shown in
Fig. \ref{fig:pd-fm-U2}.
We find that in the FM phase, the system is in a half-metallic state but is not topological, in contrast to the finite $U_d$ case shown in 
Fig. \ref{fig:phase-fm}.

\begin{figure}[H]
\centering        
\includegraphics[width=0.8\hsize,clip,bb=99 430 469 732]{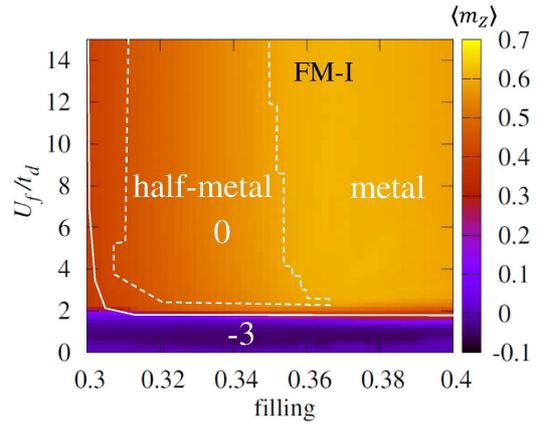}
\caption{
(Color) Phase diagram of mirror-selective topological insulator around quarter filling for $U_d=2$.
The black line separates the topological phase and trivial phase, while the white line separates the half-metallic phase and the metallic phase.
In the half-metallic region, the sector of $M_z=-i$ is an insulator and the other sector is a metal. The color plot shows the strength of the magnetization.
}      
\label{fig:pd-fm-U2}  
\end{figure}
\bibliography{68707}

\begin{thebibliography}{10}

\bibitem{Kane_Mele_QSH_05}
C.~L. Kane and E.~J. Mele: Phys. Rev. Lett. {\bfseries 95} (2005) 226801.

\bibitem{Kane_Mele_Z2_05}
C.~L. Kane and E.~J. Mele: Phys. Rev. Lett. {\bfseries 95} (2005) 146802.

\bibitem{Liang_Fu_11}
L.~Fu: Phys. Rev. Lett. {\bfseries 106} (2011) 106802.

\bibitem{Hsieh_12}
T.~H. Hsieh, H.~Lin, J.~Liu, W.~Duan, A.~Bansil, and L.~Fu: Nature
  communications {\bfseries 3} (2012) 982.

\bibitem{Tanaka_Ren_12}
Y.~Tanaka, Z.~Ren, T.~Sato, K.~Nakayama, S.~Souma, T.~Takahashi, K.~Segawa, and
  Y.~Ando: Nature Physics {\bfseries 8} (2012) 800.

\bibitem{Dziawa_Kowalski_12}
P.~Dziawa, B.~J. Kowalski, K.~Dybko, R.~Buczko, A.~Szczerbakow, M.~Szot,
  E.~{\L}usakowska, T.~Balasubramanian, B.~M. Wojek, M.~H. Bastian, M.~H.
  Berntsen, O.~Tjernberg, and T.~Story: Nature materials {\bfseries 11} (2012)
  1023.

\bibitem{Kitaev_01}
A.~Y. Kitaev: Physics-Uspekhi {\bfseries 44} (2001) 131.

\bibitem{Mourik_12}
V.~Mourik, K.~Zuo, S.~M. Frolov, S.~R. Plissard, E.~P. A.~M. Bakkers, and L.~P.
  Kouwenhoven: Science {\bfseries 336} (2012) 1003.

\bibitem{Rokhinson_12}
L.~P. Rokhinson, X.~Liu, and J.~K. Furdyna: Nature Physics {\bfseries 8} (2012)
  795.

\bibitem{Das_12}
A.~Das, Y.~Ronen, Y.~Most, Y.~Oreg, M.~Heiblum, and H.~Shtrikman: Nature
  Physics {\bfseries 8} (2012) 887.

\bibitem{Yoshida_Sigrist_15}
T.~Yoshida, M.~Sigrist, and Y.~Yanase: Phys. Rev. Lett. {\bfseries 115} (2015)
  027001.

\bibitem{Daido_16}
A.~Daido and Y.~Yanase: Phys. Rev. B {\bfseries 94} (2016) 054519.

\bibitem{Murakami_07}
S.~Murakami: New Journal of Physics {\bfseries 9} (2007) 356.

\bibitem{Wan_11}
X.~Wan, A.~M. Turner, A.~Vishwanath, and S.~Y. Savrasov: Phys. Rev. B
  {\bfseries 83} (2011) 205101.

\bibitem{Burkov_11}
A.~A. Burkov and L.~Balents: Phys. Rev. Lett. {\bfseries 107} (2011) 127205.

\bibitem{Weng_15}
H.~Weng, C.~Fang, Z.~Fang, B.~A. Bernevig, and X.~Dai: Phys. Rev. X {\bfseries
  5} (2015) 011029.

\bibitem{Xu_15}
S.-Y. Xu, I.~Belopolski, N.~Alidoust, M.~Neupane, G.~Bian, C.~Zhang, R.~Sankar,
  G.~Chang, Z.~Yuan, C.-C. Lee, S.-M. Huang, H.~Zheng, J.~Ma, D.~S. Sanchez,
  B.~Wang, A.~Bansil, F.~Chou, P.~P. Shibayev, H.~Lin, S.~Jia, and M.~Z. Hasan:
  Science {\bfseries 349} (2015) 613.

\bibitem{Lv_15}
B.~Q. Lv, H.~M. Weng, B.~B. Fu, X.~P. Wang, H.~Miao, J.~Ma, P.~Richard, X.~C.
  Huang, L.~X. Zhao, G.~F. Chen, Z.~Fang, X.~Dai, T.~Qian, and H.~Ding: Phys.
  Rev. X {\bfseries 5} (2015) 031013.

\bibitem{Dzero_10}
M.~Dzero, K.~Sun, V.~Galitski, and P.~Coleman: Phys. Rev. Lett. {\bfseries 104}
  (2010) 106408.

\bibitem{Dzero_16}
M.~Dzero, J.~Xia, V.~Galitski, and P.~Coleman: Annual Review of Condensed
  Matter Physics {\bfseries 7} (2016) 249.

\bibitem{Neupane_13}
M.~Neupane, N.~Alidoust, S.-Y. Xu, T.~Kondo, Y.~Ishida, D.-J. Kim, C.~Liu,
  I.~Belopolski, Y.~J. Jo, T.-R. Chang, H.-T. Jeng, T.~Durakiewicz, L.~Balicas,
  H.~Lin, A.~Bansil, S.~Shin, Z.~Fisk, and M.~Z. Hasan: Nature communications
  {\bfseries 4} (2013) 2991.

\bibitem{Xu_14}
N.~Xu, P.~K. Biswas, J.~H. Dil, R.~S. Dhaka, G.~Landolt, S.~Muff, C.~E. Matt,
  X.~Shi, N.~Plumb, M.~Radovi{\'c}, E.~Pomjakushina, K.~Conder, A.~Amato, S.~V.
  Borisenko, R.~Yu, H.-M. Weng, Z.~Fang, X.~Dai, J.~Mesot, H.~Ding, and M.~Shi:
  Nature communications {\bfseries 5} (2014) 4566.

\bibitem{Hohenadler_11}
M.~Hohenadler, T.~C. Lang, and F.~F. Assaad: Phys. Rev. Lett. {\bfseries 106}
  (2011) 100403.

\bibitem{Yoshida_Fujimoto_Kawakami_12}
T.~Yoshida, S.~Fujimoto, and N.~Kawakami: Phys. Rev. B {\bfseries 85} (2012)
  125113.

\bibitem{Tada_12}
Y.~Tada, R.~Peters, M.~Oshikawa, A.~Koga, N.~Kawakami, and S.~Fujimoto: Phys.
  Rev. B {\bfseries 85} (2012) 165138.

\bibitem{Shitade_09}
A.~Shitade, H.~Katsura, J.~Kune\ifmmode~\check{s}\else \v{s}\fi{}, X.-L. Qi,
  S.-C. Zhang, and N.~Nagaosa: Phys. Rev. Lett. {\bfseries 102} (2009) 256403.

\bibitem{Chadov_10}
S.~Chadov, X.~Qi, J.~K{\"u}bler, G.~H. Fecher, C.~Felser, and S.~C. Zhang: Nat.
  Mater. {\bfseries 9} (2010) 541.

\bibitem{Lin_10}
H.~Lin, L.~A. Wray, Y.~Xia, S.~Xu, S.~Jia, R.~J. Cava, A.~Bansil, and M.~Z.
  Hasan: Nat. Mater. {\bfseries 9} (2010) 546.

\bibitem{Takimoto_11}
T.~Takimoto: Journal of the Physical Society of Japan {\bfseries 80} (2011)
  123710.

\bibitem{Yan_12}
B.~Yan, L.~M\"uchler, X.-L. Qi, S.-C. Zhang, and C.~Felser: Phys. Rev. B
  {\bfseries 85} (2012) 165125.

\bibitem{Zhang_13}
X.~Zhang, N.~P. Butch, P.~Syers, S.~Ziemak, R.~L. Greene, and J.~Paglione:
  Phys. Rev. X {\bfseries 3} (2013) 011011.

\bibitem{Lu_13}
F.~Lu, J.~Zhao, H.~Weng, Z.~Fang, and X.~Dai: Phys. Rev. Lett. {\bfseries 110}
  (2013) 096401.

\bibitem{Hsieh_14}
T.~H. Hsieh, J.~Liu, and L.~Fu: Phys. Rev. B {\bfseries 90} (2014) 081112.

\bibitem{Kargarian_13}
M.~Kargarian and G.~A. Fiete: Phys. Rev. Lett. {\bfseries 110} (2013) 156403.

\bibitem{Weng_14}
H.~Weng, J.~Zhao, Z.~Wang, Z.~Fang, and X.~Dai: Phys. Rev. Lett. {\bfseries
  112} (2014) 016403.

\bibitem{Pesin_Balents_10}
D.~Pesin and L.~Balents: Nat. Phys. {\bfseries 6} (2010) 376.

\bibitem{Yoshida_Peters_Fujimoto_Kawakami_14}
T.~Yoshida, R.~Peters, S.~Fujimoto, and N.~Kawakami: Phys. Rev. Lett.
  {\bfseries 112} (2014) 196404.

\bibitem{Yoshida_Kawakami_16}
T.~Yoshida and N.~Kawakami: Phys. Rev. B {\bfseries 94} (2016) 085149.

\bibitem{Wu_16}
H.-Q. Wu, Y.-Y. He, Y.-Z. You, T.~Yoshida, N.~Kawakami, C.~Xu, Z.~Y. Meng, and
  Z.-Y. Lu: Phys. Rev. B {\bfseries 94} (2016) 165121.

\bibitem{Fidkowski_10}
L.~Fidkowski and A.~Kitaev: Phys. Rev. B {\bfseries 81} (2010) 134509.

\bibitem{Fidkowski_11}
L.~Fidkowski and A.~Kitaev: Phys. Rev. B {\bfseries 83} (2011) 075103.

\bibitem{Turner_11}
A.~M. Turner, F.~Pollmann, and E.~Berg: Phys. Rev. B {\bfseries 83} (2011)
  075102.

\bibitem{Ryu_Zhang_12}
S.~Ryu and S.-C. Zhang: Phys. Rev. B {\bfseries 85} (2012) 245132.

\bibitem{Yao_Ryu_13}
H.~Yao and S.~Ryu: Phys. Rev. B {\bfseries 88} (2013) 064507.

\bibitem{Qi_13}
X.-L. Qi: New J. Phys. {\bfseries 15} (2013) 065002.

\bibitem{Lu_Vishwanath_12}
Y.-M. Lu and A.~Vishwanath: Phys. Rev. B {\bfseries 86} (2012) 125119.

\bibitem{Levin_12}
M.~Levin and A.~Stern: Phys. Rev. B {\bfseries 86} (2012) 115131.

\bibitem{Fidkowski_13}
L.~Fidkowski, X.~Chen, and A.~Vishwanath: Phys. Rev. X {\bfseries 3} (2013)
  041016.

\bibitem{Gu_14}
Z.-C. Gu and X.-G. Wen: Phys. Rev. B {\bfseries 90} (2014) 115141.

\bibitem{Hsieh_Chang_14}
C.-T. Hsieh, T.~Morimoto, and S.~Ryu: Phys. Rev. B {\bfseries 90} (2014)
  245111.

\bibitem{Isobe_Fu_15}
H.~Isobe and L.~Fu: Phys. Rev. B {\bfseries 92} (2015) 081304.

\bibitem{Yoshida_Furusaki_15}
T.~Yoshida and A.~Furusaki: Phys. Rev. B {\bfseries 92} (2015) 085114.

\bibitem{Wang_Potter_Senthil_14}
C.~Wang, A.~C. Potter, and T.~Senthil: Science {\bfseries 343} (2014) 629.

\bibitem{You_Cenke_14}
Y.-Z. You and C.~Xu: Phys. Rev. B {\bfseries 90} (2014) 245120.

\bibitem{Wang_Senthil_14}
C.~Wang and T.~Senthil: Phys. Rev. B {\bfseries 89} (2014) 195124.

\bibitem{Morimoto_15}
T.~Morimoto, A.~Furusaki, and C.~Mudry: Phys. Rev. B {\bfseries 92} (2015)
  125104.

\bibitem{Yoshida_Daido_17}
T.~Yoshida, A.~Daido, Y.~Yanase, and N.~Kawakami: Phys. Rev. Lett. {\bfseries
  118} (2017) 147001.

\bibitem{Yoshida_Danshita_17}
T.~Yoshida, I.~Danshita, R.~Peters, and N.~Kawakami: arXiv preprint
  arXiv:1711.09538 .

\bibitem{Mong_10}
R.~S.~K. Mong, A.~M. Essin, and J.~E. Moore: Phys. Rev. B {\bfseries 81} (2010)
  245209.

\bibitem{Essin_Gurarie_12}
A.~M. Essin and V.~Gurarie: Phys. Rev. B {\bfseries 85} (2012) 195116.

\bibitem{Okamoto_14}
S.~Okamoto, W.~Zhu, Y.~Nomura, R.~Arita, D.~Xiao, and N.~Nagaosa: Phys. Rev. B
  {\bfseries 89} (2014) 195121.

\bibitem{Guo_11}
H.~Guo, S.~Feng, and S.-Q. Shen: Phys. Rev. B {\bfseries 83} (2011) 045114.

\bibitem{He_11}
J.~He, Y.-H. Zong, S.-P. Kou, Y.~Liang, and S.~Feng: Phys. Rev. B {\bfseries
  84} (2011) 035127.

\bibitem{Yoshida_13}
T.~Yoshida, R.~Peters, S.~Fujimoto, and N.~Kawakami: Phys. Rev. B {\bfseries
  87} (2013) 085134.

\bibitem{He_12}
J.~He, B.~Wang, and S.-P. Kou: Phys. Rev. B {\bfseries 86} (2012) 235146.

\bibitem{Yoshida_SSTI_13}
T.~Yoshida, R.~Peters, S.~Fujimoto, and N.~Kawakami: Phys. Rev. B {\bfseries
  87} (2013) 165109.

\bibitem{Rachel_10}
S.~Rachel and K.~Le~Hur: Phys. Rev. B {\bfseries 82} (2010) 075106.

\bibitem{Varney_10}
C.~N. Varney, K.~Sun, M.~Rigol, and V.~Galitski: Phys. Rev. B {\bfseries 82}
  (2010) 115125.

\bibitem{Yamaji_11}
Y.~Yamaji and M.~Imada: Phys. Rev. B {\bfseries 83} (2011) 205122.

\bibitem{Zheng_11}
D.~Zheng, G.-M. Zhang, and C.~Wu: Phys. Rev. B {\bfseries 84} (2011) 205121.

\bibitem{Griset_12}
C.~Griset and C.~Xu: Phys. Rev. B {\bfseries 85} (2012) 045123.

\bibitem{Wu_12}
W.~Wu, S.~Rachel, W.-M. Liu, and K.~Le~Hur: Phys. Rev. B {\bfseries 85} (2012)
  205102.

\bibitem{Yu_11}
S.-L. Yu, X.~C. Xie, and J.-X. Li: Phys. Rev. Lett. {\bfseries 107} (2011)
  010401.

\bibitem{ten-fold_way_08}
A.~P. Schnyder, S.~Ryu, A.~Furusaki, and A.~W.~W. Ludwig: Phys. Rev. B
  {\bfseries 78} (2008) 195125.

\bibitem{ten-fold_way_09}
A.~Kitaev: AIP Conference Proceedings {\bfseries 1134} (2009) 22.

\bibitem{ten-fold_way_10}
S.~Ryu, A.~P. Schnyder, A.~Furusaki, and A.~W.~W. Ludwig: New Journal of
  Physics {\bfseries 12} (2010) 065010.

\bibitem{Slager_13}
R.-J. Slager, A.~Mesaros, V.~Juri{\v{c}}i{\'c}, and J.~Zaanen: Nature Physics
  {\bfseries 9} (2013) 98.

\bibitem{Slager_17}
J.~Kruthoff, J.~de~Boer, J.~van Wezel, C.~L. Kane, and R.-J. Slager: Phys. Rev.
  X {\bfseries 7} (2017) 041069.

\bibitem{Shiozaki_nonsymm_16}
K.~Shiozaki, M.~Sato, and K.~Gomi: Phys. Rev. B {\bfseries 93} (2016) 195413.

\bibitem{Beach_Assaad_08}
K.~S.~D. Beach and F.~F. Assaad: Phys. Rev. B {\bfseries 77} (2008) 205123.

\bibitem{Peters_12}
R.~Peters, N.~Kawakami, and T.~Pruschke: Phys. Rev. Lett. {\bfseries 108}
  (2012) 086402.

\bibitem{Howczak_Spalek_12}
O.~Howczak and J.~Spalek: Journal of Physics: Condensed Matter {\bfseries 24}
  (2012) 205602.

\bibitem{Peters_Kawakami_12}
R.~Peters and N.~Kawakami: Phys. Rev. B {\bfseries 86} (2012) 165107.

\bibitem{Denis_Rok_13}
D.~Gole\ifmmode~\check{z}\else \v{z}\fi{} and R.~\ifmmode~\check{Z}\else
  \v{Z}\fi{}itko: Phys. Rev. B {\bfseries 88} (2013) 054431.

\bibitem{Legner_Sigrist_14}
M.~Legner, A.~R\"uegg, and M.~Sigrist: Phys. Rev. B {\bfseries 89} (2014)
  085110.

\bibitem{Legner_Sigrist_15}
M.~Legner, A.~R\"uegg, and M.~Sigrist: Phys. Rev. Lett. {\bfseries 115} (2015)
  156405.

\bibitem{Tran_12}
M.-T. Tran, T.~Takimoto, and K.-S. Kim: Phys. Rev. B {\bfseries 85} (2012)
  125128.

\bibitem{Alexandrov_13}
V.~Alexandrov, M.~Dzero, and P.~Coleman: Phys. Rev. Lett. {\bfseries 111}
  (2013) 226403.

\bibitem{Fukui_Hatsugai_05}
T.~Fukui, Y.~Hatsugai, and H.~Suzuki: Journal of the Physical Society of Japan
  {\bfseries 74} (2005) 1674.

\bibitem{Rui-Xing_Zhang_16}
R.-X. Zhang, C.~Xu, and C.-X. Liu: Phys. Rev. B {\bfseries 94} (2016) 235128.

\bibitem{Volovik_03}
G.~E. Volovik: {\em The universe in a helium droplet} (Oxford University Press
  on Demand, 2003), Vol. 117.

\bibitem{Gurarie_11}
V.~Gurarie: Phys. Rev. B {\bfseries 83} (2011) 085426.

\bibitem{Zhong_Wang_12}
Z.~Wang and S.-C. Zhang: Phys. Rev. X {\bfseries 2} (2012) 031008.

\bibitem{Essin_Gurarie_11}
A.~M. Essin and V.~Gurarie: Phys. Rev. B {\bfseries 84} (2011) 125132.

\bibitem{Zhong_13}
Z.~Wang and B.~Yan: Journal of Physics: Condensed Matter {\bfseries 25} (2013)
  155601.

\bibitem{Peters_18}
R.~Peters, T.~Yoshida, and N.~Kawakami: arXiv preprint arXiv:1804.04802 .

\end{thebibliography}

\end{document}